\documentclass[twocolumn,letter]{jpsj2} 
\title{Charge order and superconductivity in a two-dimensional triangular lattice at $n=2/3$}

\author{Hiroshi \textsc{Watanabe}\thanks{E-mail address: hwatanabe@hosi.phys.s.u-tokyo.ac.jp} and Masao \textsc{Ogata}}

\inst{Department of Physics, University of Tokyo, Hongo, Bunkyo-ku, Tokyo 113-0033 Japan}

\abst{To investigate the possibility of charge order and superconductivity in a doped two-dimensional triangular lattice,
we study the extended Hubbard model with variational Monte Carlo method. At $n=2/3$, a commensurate filling for
a triangular lattice, it is shown that the nearest-neighbor Coulomb interaction $V$ induces honeycomb-type charge 
order and antiferromagnetic spin order at $U\gtrsim10t$. We also discuss the possibility of superconductivity induced 
by charge fluctuation and the relation to the superconductivity in Na$_{0.35}$CoO$_2\cdot 1.3$H$_2$O and 
$\theta$-type organic condoctors.}

\kword{charge order, charge fluctuation, superconductivity, Na$_{0.35}$CoO$_2\cdot 1.3$H$_2$O, 
          $\theta$-type organic conductor,
           variational Monte Carlo method}

\begin{document}
\maketitle

Since the discovery of high-$T_c$ cuprates, it have been clarified that superconductivity and magnetism are closely
connected with each other in strongly correlated electron systems. Hubbard model and $t$-$J$ model in a two-dimensional 
square lattice have successfully described the relation between them and showed that antiferromagnetic spin fluctuation 
induces $d_{x^2-y^2}$-wave superconductivity in doped Mott insulators.

In a frustrated system like a two-dimensional triangular lattice, on the other hand, the relation between superconductivity
and magnetism have not been clarified yet. Generally, frustration suppresses long-range magnetic orders and induces large 
fluctuation. Anderson first proposed a resonating valence bond (RVB) state for the ground state of an antiferromagnetic
two-dimensional triangular lattice, as a notion substituted for the Fermi liquid.~\cite{Anderson} Since then, study of frustrated spin 
systems has been greatly improved and various interesting phenomena have been clarified. However, when some carriers
are doped into the frustrated lattices, the character of electron systems has not been clarified extensively. 
Actually, materials considered as doped frustrated systems like $\theta$-type organic conductors~\cite{HMori} and
Na$_x$CoO$_2\cdot y$H$_2$O~\cite{Takada,Foo} show various magnetic and charge orders, as well as unconventional 
superconductivity.

In a system away from half filling, carriers obtain a large mobility and the charge fluctuation is enhanced by strong 
electron correlation. Then, the charge degrees of freedom become important as well as the spin degrees of freedom.
To deal with the charge fluctuation correctly, we have to take into account the long-range Coulomb interaction
which is not contained in the Hubbard model or $t$-$J$ model. An extended Hubbard model with on-site Coulomb interaction
$U$ and nearest-neighbor one $V$ is an approriate model to treat this problem. The competition between $U$ and $V$
is considered to induce various states. For example, in a two-dimensional square lattice with quarter-filled band, 
previous theoretical studies show that a checkerboard-type charge order (CO) becomes stable at large $V$
region~\cite{McKenzie} and $d_{xy}$-wave superconductivity is stabilized at intermediate $V$ region.~\cite{Merino, Kobayashi}

\begin{figure}[t]
\begin{center}
\includegraphics[width=4cm]{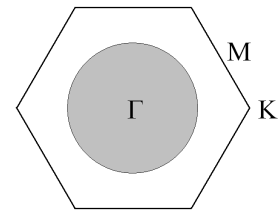}
\end{center}
\caption{First Brillouin zone and Fermi surface used in this paper.}
\label{f1}
\end{figure}

In this letter, we study an extended Hubbard model in a two-dimensional triangular lattice with variational Monte Carlo
method having the Na$_x$CoO$_2$ system in mind. Although the multi-orbital effect may be important for 
superconductivity~\cite{Mochizuki, Yanase}, we focus on the single-band model~\cite{Baskaran1, Kumar, Wang, Ogata,
Kuroki, Koretsune} with charge fluctuation.\cite{Baskaran2, Motrunich1, Tanaka}

We show that the competition between $U$ and $V$ induces the
two types of CO states and discuss the possibility of superconductivity induced by charge fluctuation.

We introduce the following extended Hubbard model in a two-dimensional triangular lattice,
\begin{equation}
 H=\sum_{\boldsymbol{k}\sigma}\epsilon_{\boldsymbol{k}}c^{\dagger}_{\boldsymbol{k}\sigma}c_{\boldsymbol{k}\sigma}
  +U\sum_{i}n_{i\uparrow}n_{i\downarrow}+V\sum_{\left<i,j\right>}n_{i}n_{j}, \label{ExtHub}
\end{equation}
where $\left<i,j\right>$ denotes the summation over the nearest-neighbor sites.
The bare energy dispersion is given by 
\begin{equation}
 \epsilon_{\boldsymbol{k}}=-2t(\cos k_x+2\cos \frac{1}{2}k_x\cos \frac{\sqrt{3}}{2}k_y). 
\end{equation}
In this non-bipartite lattice, there is no electron-hole symmetry and the sign of $t$ is crucial. 
The sign of $t$ should be determined so as to reproduces the shape of the Fermi surface obtained from experiments
or band calculations for each material under consideration. For the superconducting material 
Na$_{0.35}$CoO$_2\cdot 1.3$H$_2$O, the choice of  $t<0$ and $n=4/3$ reproduces the large Fermi surface around
$\Gamma$ point calculated by the LDA calculation~\cite{Singh} or observed by ARPES experiments.~\cite{Hasan,Yang}
In the following, we use $t>0$ and $n=2/3$ by using the electron-hole transformation as shown in Fig.~\ref{f1}.
To study the ground state, we use variational Monte Carlo (VMC) method. We introduce the Jastrow-type
trial wave function for the extended Hubbard model,
\begin{equation}
 \left|\Psi \right> = P_W P_G \left|\Phi \right>, 
\end{equation}
where
\begin{align}
 P_G &= \prod_{i}\left(1-(1-g)n_{i\uparrow}n_{i\downarrow}\right), \\
 P_W &= w^{\sum_{\left<i,j\right>}n_in_j}.
\end{align} 
$P_G$ is a Gutzwiller projection operator which reduces the probability of double occupancy of electrons at the
same site. $P_W$ is a nearest-neighbor projection operator which controls the weight of the electron configuration
at the nearest-neighbor sites. 
This kind of projection operator is successfully used in the square lattice.~\cite{Yokoyama1, Yokoyama2}
$g$ and $w$ are variational parameters with $0\leq g, w\leq 1$. 
If we set $g=w=1$, $\left|\Psi \right>$ is reduced to $\left|\Phi \right>$ which is constructed from a mean-field 
solution of a certain Hamiltonian and its explicit form is discussed shortly. 

\begin{figure}[t]
\begin{center}
\includegraphics[width=8.5cm]{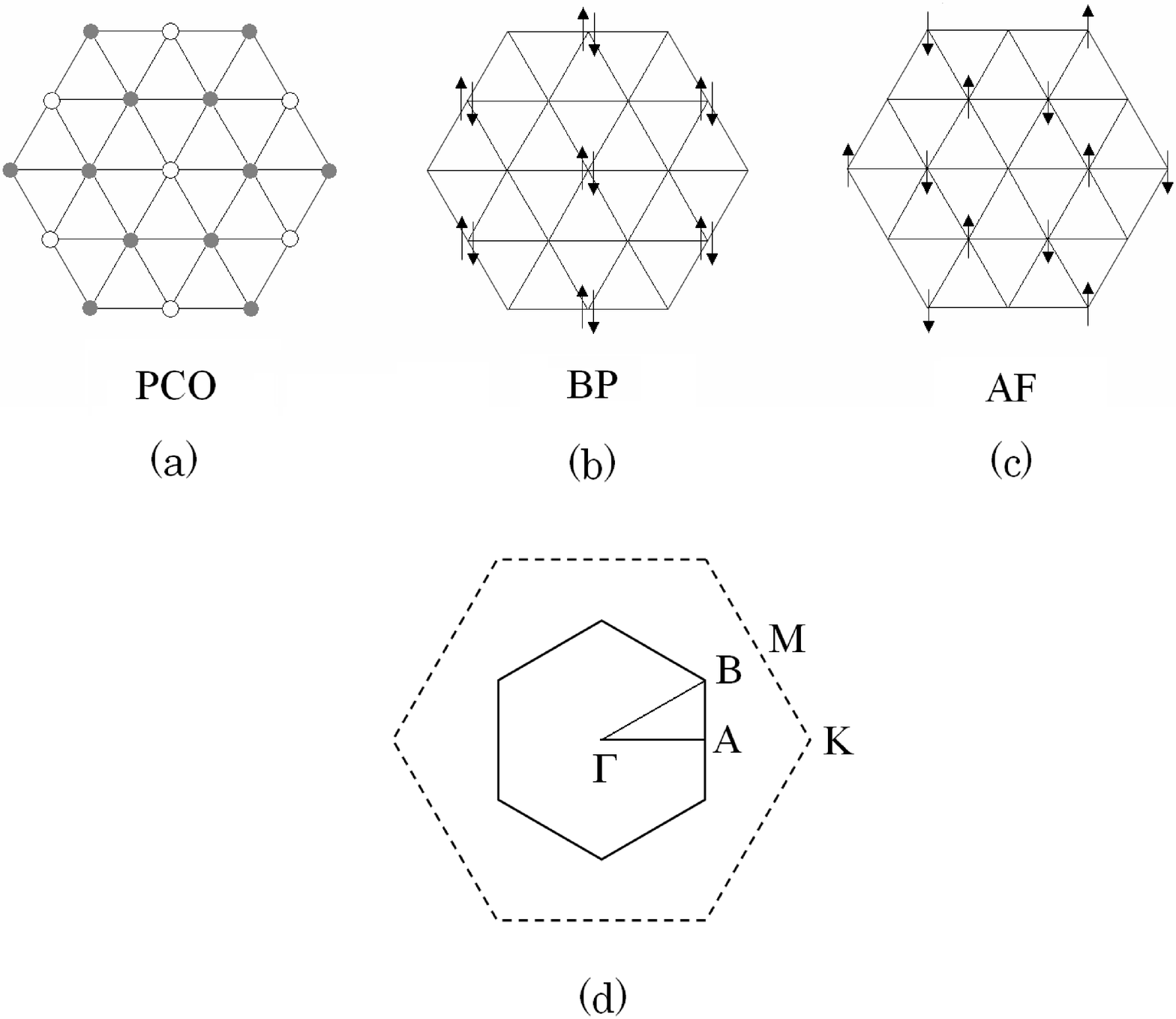}
\end{center}
\caption{Three different charge and spin ordered states in a two-dimensional triangular lattice for $n=2/3$. 
(a) Paramagnetic CO where solid (open) circles represent preferred B and C (avoided A) sublattices. (b) Bipolaronic state
where electrons prefer A sublattice in spite of double occupancy. (c) AF - Antiferromagnetic state with honeycomb-type
charge order. (d) The folded first Brillouin zone for a honeycomb-type charge order. Dashed line represents the original
first Brillouin zone.}
\label{f2}
\end{figure}

In a two-dimensional triangular lattice at $n=2/3$, several honeycomb-type states as shown in Fig.~\ref{f2} are 
commensurate charge-ordered states. The competition between $U$ and $V$ is expected to induce some charge-ordered states. 
In the following, we denote the three different states as paramagnetic charge order (PCO), 
bipolaronic charge order (BP), and antiferromagnetic charge order (AF) as shown in 
Fig.~\ref{f2}(a)-(c). Trial wave functions corresponding to
these states are obtained from the solution of the mean-fileld Hamiltonian
\begin{equation}
 \begin{split}
 H_{\sigma}&=\sum_{\boldsymbol{k}\sigma}\epsilon_{\boldsymbol{k}}c^{\dagger}_{\boldsymbol{k}\sigma}c_{\boldsymbol{k}\sigma}
  +2\Delta_{\mathrm{c}}\sum_{i}\cos(\boldsymbol{Q}\cdot\boldsymbol{r}_i)n_i \\
  &-2\Delta_{\mathrm{s}}\sum_{i}\cos\left(\boldsymbol{Q}\cdot\boldsymbol{r}_i+\sigma\frac{2\pi}{3}\right)n_{i\sigma}, \label{Ham}
 \end{split}
\end{equation}
where $\Delta_{\mathrm{c}}$ is an order parameter of charge order, $\Delta_{\mathrm{s}}$ is an order parameter of spin 
order and $\boldsymbol{Q}=(\frac{4\pi}{3},0)$ is an ordering wave vector. PCO is a state in which electrons prefer B and C sublattices and
avoid A sublattice, and there is no spin order. Thus we can choose $\Delta_{\mathrm{c}}>0, \Delta_{\mathrm{s}}=0$. BP is a bipolaronic
state in which electrons prefer A sublattice and avoid B and C sublattices with double occupancy, i.e., $\Delta_{\mathrm{c}}<0, 
\Delta_{\mathrm{s}}=0$. Finally, AF is a state that charge order and antiferromagnetic spin order coexist~\cite{Ueda}
($\Delta_{\mathrm{c}}>0,\Delta_{\mathrm{s}}>0$). If we set $\Delta_{\mathrm{c}}=\Delta_{\mathrm{s}}=0$, the state 
reduces to paramagnetic metal (PM).

\begin{figure}[t]
\begin{center}
\includegraphics[width=6cm]{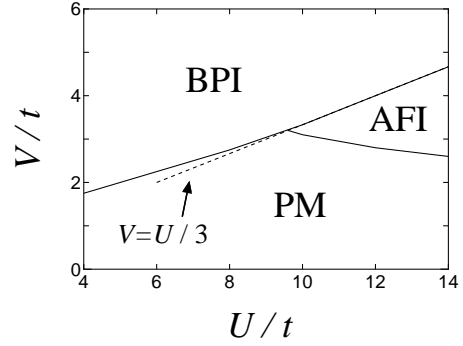}
\end{center}
\caption{Phase diagram in a $U-V$ plane for $n=128/192=2/3$. Solid lines represent the first-order transition 
boundaries. The boundary between BPI and AFI is well fitted by the dotted line of $V=U/3$.}
\label{f3}
\end{figure}

For the trial wave fuction of superconductivity, we employ the BCS type wave function,
\begin{equation}
 \left|\Phi\right>=\prod_{\boldsymbol{k}}\left(u_{\boldsymbol{k}}+v_{\boldsymbol{k}}c^{\dagger}_{\boldsymbol{k}\uparrow}
 c^{\dagger}_{-\boldsymbol{k}\downarrow}\right)\left|0\right>.
\end{equation}
The form of this wave function is not convenient for the VMC method because the number of electrons are not fixed.
To obtain the wave function with a fixed number of electrons, we extract the term imvolving $N_{\mathrm{e}}$ electrons
as,\cite{Gros} 
\begin{equation}
\left|\Phi\right>\propto\left(\sum_{i,j}a_{ij}c^{\dagger}_{i\uparrow}c^{\dagger}_{j\downarrow}\right)^{N_{\mathrm{e}}/2}
                   \left|0\right>
\end{equation}
where
\begin{align}
 a_{ij}&=\frac{1}{\sqrt{N_{\mathrm{e}}}}\sum_{\boldsymbol{k}}\frac{v_{\boldsymbol{k}}}{u_{\boldsymbol{k}}}\mathrm{e}^{\mathrm{i
}\boldsymbol{k}\cdot(\boldsymbol{r}_i-\boldsymbol{r}_j)} \notag \\
       &=\frac{1}{\sqrt{N_{\mathrm{e}}}}\sum_{\boldsymbol{k}}\frac{\Delta_{\boldsymbol{k}}}{\xi_{\boldsymbol{k}}
        +\sqrt{\xi^2_{\boldsymbol{k}}+\left|\Delta_{\boldsymbol{k}}\right|^2}}\mathrm{e}^{\mathrm{i}\boldsymbol{k}
        \cdot(\boldsymbol{r}_i-\boldsymbol{r}_j)}.
\end{align}
The matrix element $a_{ij}$ reflects the symmetry of Cooper pairs: for singlet pairing, $a_{ji}=a_{ij}$ and for triplet,
$a_{ji}=-a_{ij}$. The possible paring symmetry of $\Delta_{\boldsymbol{k}}$, the gap function of superconductivity, is 
determined by group theory. The six candidates are $s,d,i$ (singlet) and $p,f,$ next-nearest-neighbor $f$ (triplet)
symmetries. 

We determine the phase diagram in a $U-V$ plane by optimizing the ground state energies of trial wave functions.
Figure~\ref{f3} shows the phase diagram in a $U-V$ plane. For large value of $V$, BPI (``I'' denotes insulator)
state is stable because the effect of $V$ overcomes $U$ and the electrons concentrate in one of the three 
sublattices in spite of double occupancies. At $U\gtrsim 10t$, there appears an AFI region between paramagnetic 
metal (PM) and BPI regions. For small $U$, the state changes as PM $\rightarrow$ BPI as $V$ increases. For large 
$U$, however, electrons try to order without loss of on-site potential energy. Then the state changes as PM 
$\rightarrow$ AFI $\rightarrow$ BPI. Since the energy loss in the Coulomb interactions in BPI and AFI are $U/3$ 
and $V$ per site, the critical value of $V\simeq U/3$ can be easily expected. 
On the other hand, we find that PCO region does not exist in this phase diagram because the calculated total energy 
of PCO is always higher than AFI. That is to say, the order parameters of charge order and spin order grow 
simultaneously.

\begin{figure}[t]
\begin{center}
\includegraphics[width=8.5cm]{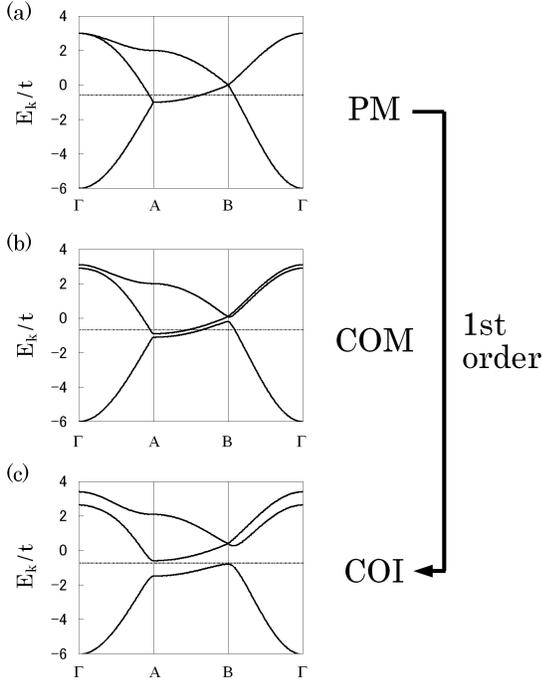}
\end{center}
\caption{Band dispersion obtained from the mean-field solution of eq.\eqref{Ham}. (a) Paramagnetic metal  
($\Delta_{\mathrm{c}}=0$), (b) charge ordered metal ($\Delta_{\mathrm{c}}=-0.1t$), (c) charge ordered insulator
($\Delta_{\mathrm{c}}=-0.5t$) which is the BPI. Horizontal lines represent the Fermi energy.}
\label{f4}
\end{figure}

Interestingly, we find that all three phase transitions (PM $\rightarrow$ BPI, PM $\rightarrow$ AFI and AFI $\rightarrow$
BPI) are first order. At $V=V_{\mathrm{c}}$, the critical value of $V$, the order parameter has a jump and the variational 
energies of the states cross. Reflecting the first order character, the physical quantities like magnetization or charge 
density change discontinuously. In order to understand the reason of this first order transition,
we show the change of the band structure in Fig.~\ref{f4}. When $\Delta_{\mathrm{c}}=0$, the system 
is a paramagnetic metal (PM, Fig.~\ref{f4}(a)) state. As $\Delta_{\mathrm{c}}$ grows, two degenerate bands along A-B line
(see Fig.~\ref{f2}(d)) begin to split. When $|\Delta_{\mathrm{c}}|<t/3$, the system shows a metallic behavior because 
there is an overlap between the upper band and lower band. We call this state as ``charge ordered metal 
(COM, Fig.~\ref{f4}(b))''. If $|\Delta_{\mathrm{c}}|>t/3$, the overlap disappears and we call this state as `` charge ordered 
insulator (COI, Fig.~\ref{f4}(c))''. PM $\rightarrow$ BPI transiton in Fig.~\ref{f3} is, that is to say, ``PM to COI'' type 
first-order transition. COM state is not realized in this condition. This result is interpreted as follows. Since $n=2/3$ is a
commensurate filling for a two-dimensional triangular lattice, the honeycomb-type charge orders like BPI or AFI are 
greatly favored. We consider that the optimal carrier doping removes 
the frustration and reduces the charge fluctuation. COM is an intermediate state with weak stability and thus it is 
suppressed in this condition. Therefore, if we slightly break the commensurability, for example, by the
introduction of a lattice anisotropy or doping, the fluctuation is enhanced again and COM state can appear with
second-order transition. Indeed, the study of the extended Hubbard model in a two-dimensional anisotropic triangular 
lattice by Seo shows the existence of COM state in some parameter region.~\cite{Seo}     

\begin{figure}[t]
\begin{center}
\includegraphics[width=8.5cm]{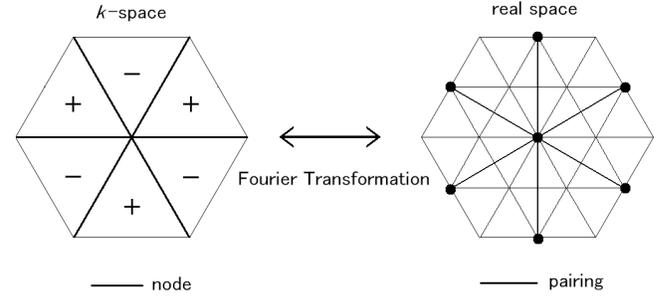}
\end{center}
\caption{Schematic representation of next-nearest-neighbor $f$-wave symmetry in a $\boldsymbol{k}$-space and a
real space. $\Delta_{\boldsymbol{k}}$ changes its sign six times in a $\boldsymbol{k}$-space and it corresponds to the
next-nearest-neighbor paring in a real space.}
\label{f5}
\end{figure}

Finally, let us discuss the possibility of superconductivity in the present case.
Our calculation shows that there is no region where a superconductivity becomes stable. This result is reasonable
at least for $V=0$ because the nesting of Fermi surface is absent. Furthermore, the van Hove singularity which is
advantageous to superconductivity is away from the Fermi surface. The third-order perturbation calculation~\cite{Ikeda}
and the one-roop renormalization-group calculation~\cite{Honerkamp} for the Hubbard model show the same results. 
Then the problem is whether the introduction of $V$ induces a superconductivity or not. Qualitatively speaking, possible 
candidate is next-nearest-neighbor (nnn) $f$-wave pairing. When $V$ is large, electrons avoid each other and nnn pairing 
is considered to be favored in the real-space representation. Schematic explanation is shown in Fig.~\ref{f5}. Indeed, we 
find that nnn $f$-wave pairing has a lower energy than the paramagnetic metal state at large value of $V$. However, 
its condensation energy is quite small compared with those of charge ordered states (BPI and AFI). 
Motrunich and Lee~\cite{Motrunich2}
also showed that the obtained $T_{\mathrm{c}}$ is much smaller than the realistic value, although the most stable pairing 
is nnn $f$-wave near the charge order instability.
We specurate that the reason why the superconductivity is not stabilized is the 
same as the case of COM. Thus, the appropriate breaking of the commensurability will induce the large fluctuation and 
then the charge order and superconductivity can compete each other. 

In the following, we propose two ways to break the commensurability and induce the superconductivity.
One is to change the electron density away from the commensurate filling. 
The RPA study of the extended Hubbard model in a two-dimensional
triangular lattice at $n=0.8$ and $1.2$ by Tanaka \textit{et al.}~\cite{Tanaka} shows that nnn $f$-wave superconductivity
is realized near the CDW instability. In this condition, CDW state is incommensurate and its stability is not so strong. 
Therefore, the charge fluctuation is enhanced near this instability and nnn $f$-wave 
superconductivity appears. Since the electron filling of Na$_{0.35}$CoO$_2\cdot 1.3$H$_2$O has not been determined 
precisely because of the existence of H$_3$O$^+$ ion, the above mechanism might be applicable to this material.

Another way is to introduce some anisotropy. Superconductivity in $\theta$-type organic 
conductors are located next to the charge order state and its origin is considered to be the charge fluctuation.
They have quasi-two-dimensional anisotropic triangular lattice with $n=1.5$ filling. For the square lattice, this filling is 
commensurate and leads to the checkerboard-type charge order. In real materials, however, the anisotropies of hopping 
or off-site Coulomb interaction break this commensurability and induces various states like several charge orders
(three kind of stripe states~\cite{Seo} and 3-fold state~\cite{TMori}) and, probably, unconventional superconductivity. 
We can relate this mechanism to the one discussed so far. The study of an anisotropic triangular lattice will interpolate 
the square- and isotropic triangular lattice and give the consensus about the charge fluctuation. It is left for the 
future problem.        

In summary, we have studied the extended Hubbard model in a two-dimensional triangular lattice and find the two types of 
charge order states (BPI and AFI) in a $U-V$ phase diagram. These states are so stable at $n=2/3$
that the superconductivity induced by charge fluctuation has no chance to appear. However, the appropriate breaking of
the commensurability can induce the charge fluctuation and the possibility of the superconductivity is expected to arise. 
We expect that this mechanism is applicable to $\theta$-type organic superconductors or 
Na$_{0.35}$CoO$_2\cdot 1.3$H$_2$O.    
\section*{Acknowledgment}
The authors thank H. Yokoyama, Y. Tanaka and M. Kaneko for useful discussions.


\begin{thebibliography}{99}
\bibitem{Ueda} We would like to thank Prof. K. Ueda for pointing out the possibility of coexistence in the early stage of
this study.
\end{thebibliography}
\end{document}